\documentclass[sigconf]{acmart}
\usepackage{balance} 
\renewcommand\footnotetextcopyrightpermission[1]{}
\usepackage[latin9]{inputenc}
\PassOptionsToPackage{normalem}{ulem}
\usepackage{ulem}
\usepackage{amsmath}
\usepackage{algorithm}
\usepackage{algpseudocode}
\usepackage{amssymb}

\let\emptyset\varnothing

\algnewcommand\algorithmicforeach{\textbf{for each}}
\algdef{S}[FOR]{ForEach}[1]{\algorithmicforeach\ #1\ \algorithmicdo}

\DeclareMathOperator*{\argmin}{argmin}

\AtBeginDocument{%
  \providecommand\BibTeX{{%
    \normalfont B\kern-0.5em{\scshape i\kern-0.25em b}\kern-0.8em\TeX}}}

\acmConference[ADF '19]{2nd KDD Workshop on Anomaly Detection in Finance 
}{August 5, 2019}{Anchorage, Alaska, USA}

\begin{document}

\title{Empirical Study on Detecting Controversy in Social Media}
\titlenote{This work is accepted and presented in the 2nd KDD Workshop on Anomaly Detection in Finance, August 5, 2019, Anchorage, Alaska, USA.}

\author{Azadeh Nematzadeh}
\authornote{The authors contributed equally to this work, listed in the alphabetical order.}
\affiliation{%
  \institution{S\&P Global}
  \streetaddress{55 Water}
  \city{New York}
  \state{NY}
  \country{USA}}
\email{azadeh.nematzadeh@spglobal.com}
\author{Grace Bang}
\authornotemark[2]
\affiliation{%
  \institution{S\&P Global}
  \streetaddress{55 Water}
  \city{New York}
  \state{NY}
  \country{USA}}
\email{grace.bang@spglobal.com}
\author{Xiaomo Liu}
\authornotemark[2]
\affiliation{%
  \institution{S\&P Global}
  \streetaddress{55 Water}
  \city{New York}
  \state{NY}
  \country{USA}}
\email{xiaomo.liu@spglobal.com}
\author{ Zhiqiang Ma}
\authornotemark[2]
\affiliation{%
  \institution{S\&P Global}
  \streetaddress{55 Water}
  \city{New York}
  \state{NY}
  \country{USA}}
\email{zhiqiang.ma@spglobal.com}
\begin{abstract}

Companies and financial investors are paying increasing attention to social consciousness in developing their corporate strategies and making investment decisions to support a sustainable economy for the future. Public discussion on incidents and events---controversies---of companies can provide valuable insights on how well the company operates with regards to social consciousness and indicate the company's overall operational capability. However, there are challenges in evaluating the degree of a company's social consciousness and environmental sustainability due to the lack of systematic data. We introduce a system that utilizes Twitter data to detect and monitor controversial events and show their impact on market volatility. In our study, controversial events are identified from clustered tweets that share the same 5W terms and sentiment polarities of these clusters. Credible news links inside the event tweets are used to validate the truth of the event. A case study on the Starbucks Philadelphia arrests shows that this method can provide the desired functionality.


  \end{abstract}

\keywords{Social Media Mining, Controversy Detection, Market Performance}
\maketitle

\section{Introduction}


The financial performance of a corporation is correlated with its social responsibility such as whether their products are environmentally friendly, manufacturing safety procedures protect against accidents, or they use child labors in its third world country factories. Consumers care about these factors when making purchasing decisions in the supermarkets and investors integrate environmental, social and governance factors, known as ESG\footnote{https://www.spglobal.com/en/capabilities/esg-evaluation}, in their investment decision-making. It has been shown that corporations financial results have a positive correlation with their sustainability business model and the ESG investment methodology can help reduce portfolio risk and generate competitive returns\footnote{https://www.forbes.com/sites/georgkell/2018/07/11/the-remarkable-rise-of-esg}. However, one barrier for ESG evaluation is the lack of relatively complete and centralized information source. Currently, ESG analysts leverage financial reports to collect the the necessary data for proper evaluation such as greenhouse gas emissions or discrimination lawsuits, but this data is inconsistent and latent. In this study, we consider social media a crowdsourcing data feed to be a new data source for this task.

Social media applications such as Twitter offer users a platform to
share and disseminate almost any content about various events such
as sports, music, and controversial events as well. The content produced
through these platforms not only facilitates the spread of information
but can also provides meaningful signals about the influence of the
events. A large number of responses to an issue on Twitter could inform the public
about the significance of an event, widen the scope of the event,
and bring more public attention inside and outside the social media
circle.

We define a controversial event for a business entity as a credible and newsworthy
incident that has the potential to impact an entity in its financial
performance and operation, for example, an incident caused by an employee
or a representative of the entity that has the potential to hurt the
trust of the public to its brand. Such an incident can demonstrate a potential gap in its risk management framework
and policy execution, and eventually hurt the interest and trust of its stakeholders'.

Controversial events trigger
a large cascade of discussion on social media platforms. The broad connectivity between people propagates their opinions into trending topics that could effect the company financially and operationally. 
In certain cases, the responsible entity can be forced to take actions,
e.g., to recall its product, which can impose a large financial burden on the
entity. For instance, in the Takata air bag scandal, the
event was discussed widely on Twitter after the New York Times published
a comprehensive article on its defective air bag products in 2014. Takata was forced to recall
nearly 50 million air bag and filed bankruptcy in June 2017.

To this end, we propose a controversial event detection
system utilizing Twitter data. We focus on controversial events which are credible and newsworthy.  Twitter data were collected on a given company and various attributes of each tweet were extracted. We verify the credibility of the event by validating the URLs appearing in tweets come from credible news sources. We utilize tweets attributes to detect events specific to the given company and the sentiment of the event to measure the controversy. Relationship between a burst of an entity controversial event and the entity market performance data was qualitatively assessed in our case study, where we found its potential impact on the equity value.

\section{Related Work}

There have been a few studies on assessing sustainability of entities.
The UN Commission on Sustainable Development (CSD) published a list
of about 140 indicators on various dimensions of sustainability \cite{CSD:2007}.
In \cite{overview:singha2009}, Singh et al. reviewed various methodologies,
indicators, and indices on sustainability assessment, which includes
environmental and social domains. All the data, on which the assessments
were conducted, mentioned in their works are processed datasets, and
some of them are collected from company annual reports and publications,
newspaper clips, and management interviews. They stated that the large
number of indicators or indices raises the need of data collection.
Our work uses the social media data as a new alternative data source
to complement the traditional data collection.

Event detection on social media has been a popular research topic
for years. Reuters Tracer \cite{Tracer:liu2016} is reported as an
application built for the journalists to detect news leads in Twitter
before the news becomes known to the public. Petrovic et al. \cite{Petrovic:2010}
presented a locality-sensitive hashing based first story detection
algorithm with a new variance reduction strategy to improve the performance.
In \cite{Event:Weng2011}, the signal of a tweet word is built with
wavelet analysis and a event is detected by clustering words with
similar signal patterns of burst. \cite{tedas:li2012} describes a
detection and analysis system named TEDAS which concentrates on Crime
and Disaster related Events (CDE). TEDAS classifies if a tweet is
a CDE tweet, predicts its geo-location if missing, and ranks and returns
important tweets when user queries in the system. TEDAS treats a tweet as an event
if the tweet qualifies, while our definition of an event is different, 
where an event is a group of tweets discussing a same theme.

\section{Controversy Detection in Social Media}

In this section, we describe the main components of our controversy detection system.

\subsection{Data collection}

The system uses Twitter's filtered streaming API to collect relevant
tweets data. The data collection pipeline accepts a comma-separated
list of phrases as filtering parameters, that the API uses to determine
which tweets will be retained from the stream. Once the system receives data from the API, it then separates postings
by companies and runs the downstream process on the separated data
streams individually.

\subsection{Feature engineering}

The data collection pipeline collects tweet postings for a given entity.
For each incoming posting, the system also stores the following attributes:
posting\_id, creation\_time, text, language, source, URLs, and hashtags.

The system parses the text attribute of each tweet.
Part-of-speech (POS) tagging and named entity recognition (NER) algorithm are applied to each tweet and terms that
are tagged as proper nouns, verbs, and entities are stored. If two proper nouns are next
to each other, the system merges them as one proper noun phrase.  Entities such as person names, organizations, locations
from tweets are the key elements in describing an event and
distinguishing it from other events, and are often used by news professionals
to describe the complete story of an event. The verbs
from POS tagging mainly represent what and why information, while
NER helps to identify where, when, and who information. They
capture the major aspects of an event, named \emph{who,
what, where, when, and why (5W)}. Besides that, 
the sentiment of each tweet is assessed too.

The system crawls the URLs in a posting and verifies whether
the link comes from one or more credible news sources. More specifically,
the system may consider the following to be examples of credible news
sources: 1) a news outlet that has, and consistently applies, journalistic
standards in its reporting or 2) an authoritative government agency
not acting in a political capacity. Determining whether a source is
a credible news source depends on the context of the event.

Based on all the extracted features, the system can build a tweet
vector, which includes the following features: tweet id, creation
time, source, hashtags, entity/proper nouns, verbs, sentiment, and news links.

\subsection{Event detection}

When a new tweet is received in the data pipeline, it either
forms a new cluster or it will be added to an existing cluster. A
new tweet will be added to an existing cluster if it is sufficiently
similar to one of the existing clusters based on its distance to the
cluster average vector. If more than one cluster is applicable, the
cluster that has the highest similarity to the new tweet is picked.
If a new tweet is not added to any existing clusters, it would form
a new cluster. A candidate event is a cluster that has at least five
tweets. Algorithm \ref{euclid} summarizes our event detection  method and the following controversy identification method.  

\subsection{Controversy identification}

An event can be controversial if the public expresses dissenting opinions,
usually associated with negative sentiments to it. The system filters
out irrelevant events and noise from the established controversial
events using the following metrics:
\begin{itemize}
\item The burstiness of an event: To detect the burstiness of an event,
the system detects the volume of tweets per time period, e.g., per
day, for the entity in question. An event is flagged when the velocity of the volume increase exceeds a threshold. 
\item Newsworthiness detection: The system counts the total number of unique
verified news links in each cluster and log that count as a newsworthiness
metric.
\item Sentiment: For each cluster, its overall sentiment score is quantified
by the mean of the sentiment scores among all tweets.
\end{itemize}
Candidate events are ranked based on these metrics, and high ranked
events are considered controversial events.

\begin{algorithm}
\caption{Outline of the controversy detection algorithm}\label{euclid}
\begin{algorithmic}[1] 

\Require $\mathcal{T}_{x} =\{ t_{1},...,t_{n} \}$ is a stream of tweets about company $x$

\Procedure{Controversy}{$\mathcal{T}_{x}$} 

\ForEach {$t \in \mathcal{T}_{x} $} \Comment(event detection)
	\State $f(t) \gets$  \Call{TweetFeature}{$t$}

	\State $\mathcal{D}_{t} \gets  \emptyset$
	\ForEach {$e_{i} \in \mathcal{E}$} \Comment{$\mathcal{E}$ current event clusters}
		\State $f(e) \gets$ \Call{ClusterFeature}{$\mathcal{E}$}
		\State $d_{t}(i) \gets$ \Call{Distance}{$t,e_{i}$}\Comment{compute distance}
		\State $\mathcal{D}_{t} \gets \{ \mathcal{D}_{t}, d_{t}(i) \}$	
	\EndFor
	
	\State $i \gets \argmin_{i} (\mathcal{D}_{t}) $  \Comment{find the closet cluster $i$}
	\If{$d_{t}(i) < D$}\Comment{$D$ is merge threshold}
		\State merge $t$ in $e_{i}$
	\Else
		\State $\mathcal{E} \gets \{  \mathcal{E}, \{t\} \}$ \Comment{$\{t\}$ is singleton cluster}
	\EndIf	
\EndFor

\State $\mathcal{E} \gets \{ e_{i} | \mathit{l}(e_{i}) > N, e_{i} \in \mathcal{E} \}$ \Comment{$N$ is min cluster size as event} 

\ForEach {$e_{i} \in \mathcal{E}$} \Comment(controversy identification)
	\State $B(e_{i})\gets$ \Call{Bustiness}{$e_{i}$} 
	\State $N(e_{i})\gets$ \Call{Newsworthiness}{$e_{i}$} 
	\ForEach {$t \in e_{i}$}
		\State $S(t) \gets$ \Call{SentimentClassify}{$t$}
		
	\EndFor
	\State $S(e_{i}) \gets$ \Call{AVG}{$S(t)$} \Comment{compute event level sentiment}
	\State $\mathcal{C} \gets S(e_{i}) < 0 \land B(e_{i}) \land N(e_{i}) $ \Comment{combined controversy score}
\EndFor	

\Return $\mathcal{C}$ \Comment{$\mathcal{C}$ is controversial events set}	 

\EndProcedure
\end{algorithmic}
\end{algorithm}

\section{Case Study - Starbucks Controversy}
\begin{figure*}[htbp]
\centering \includegraphics[width=1\textwidth]{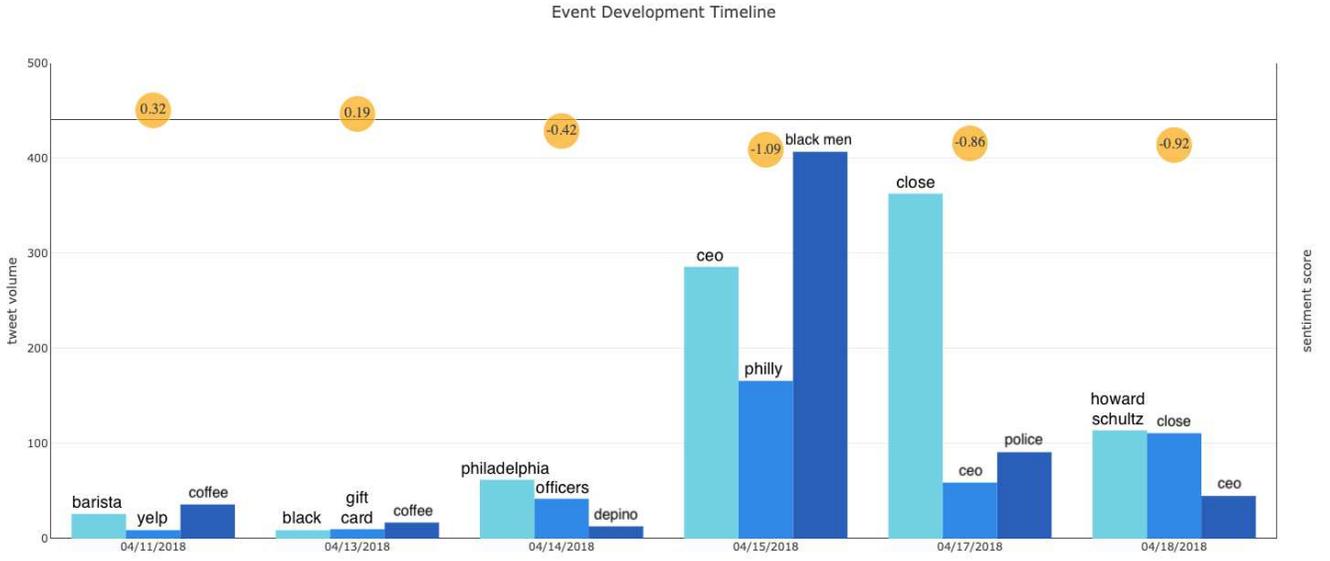} \caption{Event clusters and the sentiment polarity score along the timeline.}
\label{fig_starbucks}
\end{figure*}
In this section, we provide a case study of our model on a Starbucks controversial
event captured in the system. We validated the event with the Wikipedia
page of Starbucks\footnote{https://en.wikipedia.org/wiki/Criticism\_of\_Starbucks\#Philadelphia\_arrests}
and the major new agencies reports. After the event was detected, its impact
was further assessed by linking to the market equity data.

On April 12th, 2018, an incident occurred in a Starbucks
in Philadelphia, PA. Two African-American men were arrested
by the police officials inside that Starbucks. It was reported that
the two were denied to access the restroom by the store staff because
they did not make any purchase. While waiting at the table,
they were told by the staff to leave as they were not making any purchase. They did not comply and thus the store manager called the police and
reported that they are trespassing. The two were arrested by the officials
but released afterwards without any pressed charges. The scene
of the arresting was posted on Twitter and quickly garnered public
attention. The video had been viewed more than three millions times
in a couple of days and the major local and national news agencies
like CNN, NPR, and NYTIMES followed the development of the story.

The public outrage originating from the social media universe swiftly
triggered a series of chain reaction in the physical world. Protesters
gathered together inside and outside the Starbucks store to demand
the manager be fired. Several days later, the CEO of the Starbucks issued a public apology for
the incident on an ABC's program and stated that he would like to meet the men
to show them compassion. To remedy the bad outcome of the event,
Starbucks closed its 8,000 stores in the U.S. on May 29th for racial-bias
training for its 175K employees. A financial settlement was also established
between the two men and Starbucks corporation. This event garnered a serious
public relations crisis for Starbucks.

Figure \ref{fig_starbucks} shows the event clusters for six days
sampled between April 10th and April 20th. Given the difficulty in showing all of
the tweets that were clustered, we use the volume of key POS tagged
words (5Ws) detected in the cluster of tweets to approximate the event content.
The keywords on the top of each bar reveal aspects of
the event cluster. This controversial Starbucks event was captured in our system on April
13th, one day after the event occurred. Prior to the event, the discussion
themes about Starbucks (clusters) on Twitter were more random and included topics such as Starbucks gift card, barista, coffee as shown on 04/11/2018.
The size of the clusters and the total volume of the tweets per
day is comparably small. The first event cluster the system detected
associates with the keyword `black', where twitter users mentioned `{[}...{]}
arrested for being Black'. After the event, the volume of the tweets
per day surged multiple times more than before and kept climbing for
about a week as the event was developing. The system clearly uncovers
the events by being able to pinpoint the clustering keywords `black
men', `philly', `CEO', `close', etc. The sentiment scores of the discussion
in the clusters for each day are shown on the top part of Figure \ref{fig_starbucks}. The sentiment
score is in a range of -2 to +2, -2 standing for very negative, 0 for neutral, and 
+2 for very positive. As the figure shows, twitter users' attitude
turned from neutral to negative post the Starbucks event occurrence. The quick turn
of sentiment polarity serves as an measurement of the event being
controversy. Through the validation of the domain of the URLs quoted in the clustered
tweets, the authentication of the event is verified. All of the elements
of this event indicate that a controversy, specifically, a social related controversy, has occurred.

\begin{figure}
\includegraphics[scale=0.35]{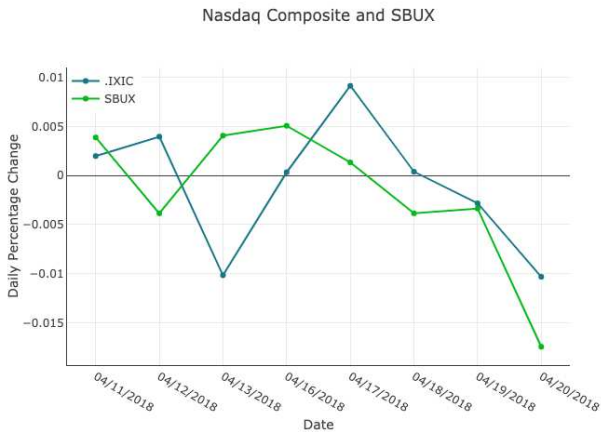}\caption{Starbucks stock price and NASDAQ index between April 11th 2018 and
April 20th 2018.}
\label{sbux_nasdaq}
\end{figure}

\begin{figure}
\centering \includegraphics[scale=0.28]{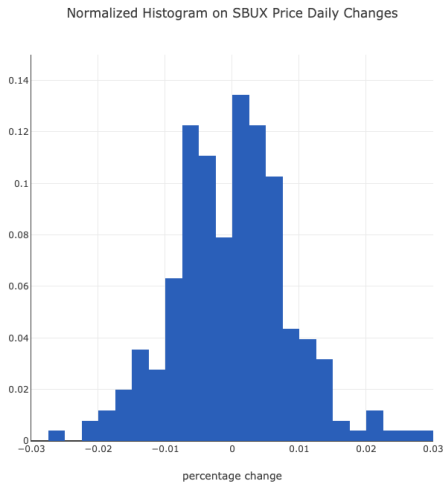}\caption{Histogram of Starbucks stock price daily changes}
\label{hist}
\end{figure}

We also did a qualitative study on the Starbucks (SBUX) stock movement
during this event. Figure \ref{sbux_nasdaq} is the daily percentage
change of SBUX and NASDAQ index between April 11th and April 20th.
SBUX did not follow the upward trend of the whole market before April
17th, and then its change on April 20th, $-1.7\%$, is quite significant from historical norms.
We collected the historical 52 week stock prices prior to
this event and calculated the daily stock price change. The distribution
of the daily price change of the previous 52 weeks is Figure \ref{hist} with a mean $\mu=4.9e-5$
and standard deviation $\sigma=0.0091$. The $1.7\%$ down almost
equals to two standard deviations below the mean. Our observation
is that plausibly, there was a negative aftereffect from the event of the notable decline in Starbucks stock price due to the
major public relations crisis.

\section{Conclusions}

We present the use of Twitter as a new data source to detect controversial events for business entities. Each tweet is represented by a vector comprising name entities and verbs mentioned in the raw tweet text. Events can be identified by grouping similar tweets in the vector space, the size and burstiness of the event, and the sentiment polarities. This system is a data-driven controversy monitoring tool that sifts through large volumes of Twitter data. It provides investors with data on key
insights on social consciousness, which allows investors to make more informed investment decisions. The direction of our future work is to: 1) develop a quantitative measure on the event impact on the equity market; 2) identify the relevance of the events to entities' operations;  3) extract post-event mitigation actions from the entities.
\balance
\bibliographystyle{ACM-Reference-Format}
\bibliography{main}
\end{document}